\documentclass[11pt,twoside,a4paper]{article}

\usepackage[utf8]{inputenc}
\usepackage{footmisc}
\usepackage{natbib}
\usepackage{graphicx}
\usepackage{amsmath}
\usepackage[hyphens]{url} 
\usepackage[disable]{todonotes}  
\let\todotmp\todo
\renewcommand{\todo}[1]{{\todotmp[nolist,color=orange!25,bordercolor=white]{#1}}}

\newcommand{\virtualrepresentation}{\textit{Virtual Representation}}

\begin{document}

\title{Virtual Representations for \\ Iterative IoT Deployment}
\author{Sebastian R. Bader, Maria Maleshkova\\
Karlsruhe Institute of Technology\\
Karlsruhe, Germany}
\date{}
\maketitle

\begin{abstract}
  A central vision of the Internet of Things is the representation of the physical world in a consistent virtual environment. Especially in the context of smart factories the connection of the different, heterogeneous production modules through a digital shop floor promises faster conversion rates, data-driven maintenance or automated machine configurations for use cases, which have not been known at design time. Nevertheless, these scenarios demand IoT representations of all participating machines and components, which requires high installation efforts and hardware adjustments.
  
  We propose an incremental process for bringing the shop floor closer to the IoT vision. Currently the majority of systems, components or parts are not yet connected with the internet and might not even provide the possibility to be technically equipped with sensors. However, those could be essential parts for a realistic digital shop floor representation. We, therefore, propose \virtualrepresentation s, which are capable of independently calculating a physical object's condition by dynamically collecting and interpreting already available data through RESTful Web APIs. The internal logic of such \virtualrepresentation \textit{s} are further adjustable at runtime, since changes to its respective physical object, its environment or updates to the resource itself should not cause any downtime.
\end{abstract}

\section{Introduction}

The ongoing digitization and interconnecting of devices of any kind through internet technologies are a prominent current trend \citep{sachs2017internet}. The Internet of Things (IoT) comprises these efforts to represent physical and virtual objects through common patterns in order to allow a digital information exchange between them. 
\citeauthor{buyya_rajkumar_internet_2013}, similarly to others, define the IoT as an ''Interconnection of sensing and actuating devices providing the ability to share information across platforms through a unified framework, developing common operating picture for enabling innovative applications. This is achieved by seamless large scale sensing, data analytics and information representation using cutting edge ubiquitous sensing and cloud computing.'' \citep{buyya_rajkumar_internet_2013}. Obviously, the regarded 'things' are restricted to physical objects, equipped with sensors or actuators, and with some kind of connection capability (referred to as 'devices'). 

We claim that the installation of sensors, the monitoring of components or the connection in digital networks is not a goal in itself but is always demand-driven. Consequently, a certain need is defined in the first place (e.g. digitally monitoring a machine's abrasion state), creating requirements on the data side. If the required data is not yet measured, additional sensors (or actuators) at the physical objects are taken into account. Only if the added value justifies this investment, the hardware is updated. Following this argumentation, the creation of an IoT setting, especially in the industrial domain, is an iterative process, where more and more things get a digital counterpart but a complete coverage of all things does not happen. In contrast to that, the key potential of the Internet of Things is not only the integration of such devices and then conducting a defined analysis but enabling previously \textit{unknown} applications. 

Especially in the context of smart factories notable efforts are conducted to bring the diverse and heterogeneous components of the shop floor to a consistent digital layer. In order to keep the integration efforts sustainable an easy to understand data and interaction model are required. The self-explaining characteristics and the high maturity level of the Semantic Web make it a suitable candidate for a future-proven solution. Such an integration layer encapsulates the complexity of underlying network implementation. Generally aspired goals are e.g. faster conversions, data-driven maintenance or automated machine configurations in order to reduce down times and, therefore, increase the overall efficiency achieved by the overlying applications. 

Even though IoT technologies support the integration process, the IoT is not only about increasing the flexibility or new insights into the behaviour of one machine but of the production line or even supply chain as a whole. Applications achieving these aims need to be deployed rapidly, if not automatically, with minimal effort. A heterogeneous landscape with a use case-driven integration approach requires continuous efforts, contradicting any potentially acquired efficiency gains. The main reason is that the configuration of interfaces is usually oriented towards scenarios relevant at the respective point in time but not for generic requirements of such later deployed systems~\cite{mariaadaptive}.
As the input needs of future applications are per se unknown at design time, a consistent and comprehensive digital reflection of the real world is more efficient and sustainable. Achieving this flexibility is the main benefit where the IoT can contribute to the manufacturing industry.

Consequently, having information about the current state of all things is crucial, e.g., the relevant machines, their respective components as well as the involved materials and products. Whereas many critical machine parts are not observable within the boundaries of an economically reasonable investment (accessibility, unreliable network access, high update costs), a continuous and complete monitoring of any system, sub-system and sub-sub-system from one day to the other is unrealistic anyway. 

We propose an incremental process for bringing the shop floor to the IoT. Many machines and devices nowadays are already (partly) IoT-ready. However, systems, modules or parts, which cannot be connected to the internet or promptly equipped with sensors are also required for a true digital shop floor. We, therefore, develop \textbf{\virtualrepresentation s}, a digital representation, which does not necessarily require a direct network connection to the physical thing. The \virtualrepresentation \ is responsible for independently calculating the object's condition by collecting and interpreting already available environment data. We present a RESTful interaction model to deploy and scale \virtualrepresentation s. Furthermore, we show how the configuration specifying a \virtualrepresentation \ can be adapted at runtime by RESTful Web APIs, since changes in the real world need to be reflected in a transparent and standardized manner. We implemented the approach based on a Linked Data Platform server, thus also enabling future extension. A demonstrating show case is also available to illustrate the concept.
We argue that the significantly lower cost of a \virtualrepresentation \ based on omitted hardware updates (if those are possible at all) as well as the newly gained opportunities to monitor, control and optimize the digital shop floor, motivates the use of  \virtualrepresentation 's.

One example application scenario is a robot gripper arm, as one unit in an industrial production line. Its motors and the supplied materials are directly observable by appropriate analytic components. In contrast, e.g. the shafts transmitting the force from the motor to the jaws and the jaws themselves cannot be monitored. Nevertheless, the abrasion state of both components might be a critical factor for a maintenance strategy. A \virtualrepresentation \ can encapsulate the necessary data collection (number and duration of loads, applied material type etc.) and the currently best performing evaluation logic. The thereby encapsulated \virtualrepresentation \ can be shifted easily, connected to other cyber-physical systems or updated as it conforms to well established Web standards. This supports the development of a sufficiently detailed and flexibly modelled smart factory, while reducing the required investment.

The paper is structured as follows: Section 2 gives a brief overview of related concepts discussed in the Web and IoT community. Section 3 further illustrates the use case, followed by the conceptional design of \virtualrepresentation s \ in Section 4. Section 5 summarizes the proposed methods and outlines future implications.

\section{Related Work}

A Thing, in the most general view, can be either a physical object, a software program, a piece of information or any other kind of identifiable resource. Within the IoT, more and more things become part of the Internet -- at the moment mainly focusing on production related objects (e.g. machines, materials etc.) or Smart Homes (smart/connected furniture). To establish a digital information exchange, a virtual resource is created in order to identify and access those objects. These resources need to be identifiable by a unique key and accessible through a commonly understood locator. Both purposes are served by URIs and their widespread usage throughout the internet makes them the identifier of choice.

The internet and, in particular, the World Wide Web offer already a well understood and widely accepted infrastructure to exchange data. Well-established Web technologies such as URIs, HTTP and hyperlinks have proven to allow easy and reliable communication mechanisms in a decentralized manner. Cloud services and on-demand Web solutions offer fast and flexible deployment of applications, a necessary requirement for a smart factory. \citeauthor{maleshkova2010investigating} analyzed the state of deployed Web APIs and RESTful services \citep{maleshkova2010investigating,bulthoff2014restful}. They conclude that a significant number of APIs lacks sufficient descriptions and at the same time miss necessary information on both the input and output data sets.
The Semantic Web on the other hand adds meaning to data objects and can reduce the mentioned integration effort. However, thereby created Web of Things does neither specify the interaction patterns of the regarded things nor does it model the intended relationship with the physical world.

\citeauthor{perera_context_2014} write about the Avatar concept as an interoperability concept between objects 'using standard protocols and technologies defined by the W3C for the Web, coupled with a distributed service-oriented mediation infrastructure' \citep{perera_context_2014}. Thus, avatars can serve as a representation for both physical and software things but do not contain formalisms to simulate unconnected objects.

'Physical entities' and 'virtual entities' are basic concepts in the domain of cyber-physical systems. \citeauthor{lee_cyber-physical_2015} introduce a general five layered architecture for cyber-physical systems. Their cyber layer includes 'cyber twins' which capture and preprocess the captured data for higher level applications \citep{lee_cyber-physical_2015}. The thereby created virtual modeling of the factory and its entities are the consistent data suppliers for the factory management. Unfortunately, they do not consider the enormous necessary effort to fully represent all machines of a production line and require to have all entities at every step and do not discuss how an environment with only partly digitized machines can benefit from their vision.

\citeauthor{glaessgen2012digital} formulate requirements on so called digital twins regarding NASA and US air force vehicles. Their goal is to simulate any perceived incident to the physical vehicle at the virtual one in order to get a higher accuracy predicting the current state of the vehicle. \citeauthor{tao2017digital} focus on the product lifecycle (design, manufacturing, service). They identify a research gap in the field of Product Lifecycle Management in the form of a disconnection between the physical object and the virtual information available during the several lifecycle stages. Their digital twin concept follows the definition of \citep{glaessgen2012digital} and focuses on information presentation in the form of a virtual objects but do not consider any manipulations of those. Therefore, their digital twin concept mainly serves as a virtual model and information container, and the thing itself as any (virtual) interaction pattern is missing.

\begin{figure}
\centering
\includegraphics[width=0.43\textwidth]{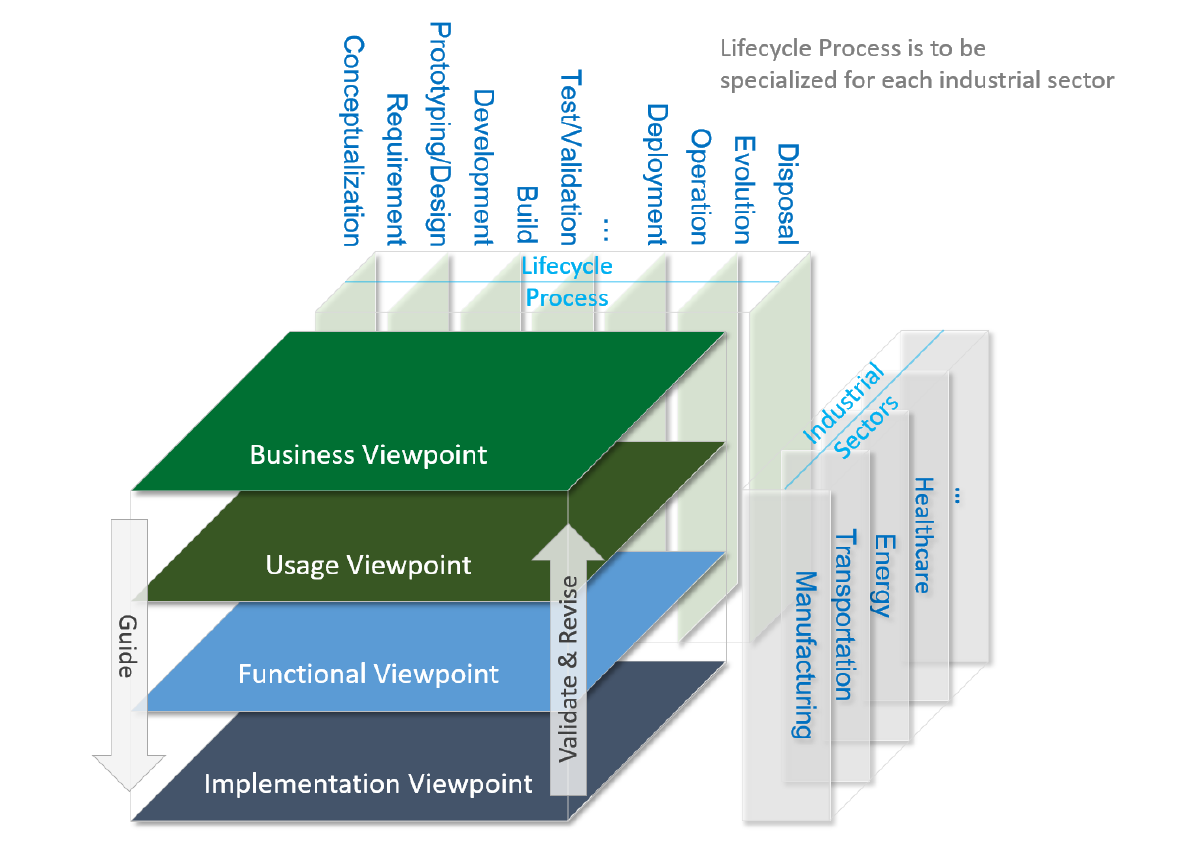}
\caption{Architectural Framework according to \protect\citep{iic_reference_architecture} }
\label{fig:IccReferenceArchitecture}
\end{figure}

The Industrial Internet Consortium aims to introduce common standards for the Industrial Internet of Things. As part of their reference architecture (Fig. \ref{fig:IccReferenceArchitecture}) the consortium discusses the trend to ''base their control decisions on the simulation model rather than a control engineer’s equation'' \citep{iic_reference_architecture}. Especially the introduced Functional Viewpoint contains the modeling of things but still only targets objects directly equipped with sensors or actuators. In addition, context information is briefly discussed but limited to the semantically described relationships of the previously mentioned entities.

Similarly to the Industrial Internet Consortium the Plattform Industrie 4.0 proposes the 'Asset Administration Shell' containing an identifier, a referent, the referent definition and a set of characteristics \citep{bedenbender_industrie_2017}. It contains basic descriptions (header) of its provided data objects and functions in the body. An 'Asset Administration Shell' can be seen as a specification of the virtual entity concept from cyber-physical systems but does also only regard connected objects. Even though an 'Asset Administration Shell' can also represent software or a Web service the authors do not discuss any strategies how their general concept also allows the introduction of unconnected things to the digital shop floor.

Edge computing \citep{hu2015mobile}, edge analytics or fog computing \citep{bonomi2012fog} proposes to conduct substantial part of computing tasks close to the physical thing at the edge of the network. The main advantages are less network traffic and response times as data processing takes place close to the data sources. Even though simulating resources close to their physical location can be seen as some kind of edge computing, the paradigm itself lacks -- to the best of our knowledge -- standardized methods describing how to implement and how to interact with the resources of interest.

The Resource Description Framework (RDF) models resources in the form of triples (subject, predicate, object). Each part of the triple can either be a resource in the form of a globally unique URI or, in case of the object position, also a sequence of characters referred to as Literals. RDF also allows several serialization formats and, therefore, serves as a flexible data format for both descriptions and productive data.
Linked Data further introduces dereferenceable identifiers (HTTP URIs) and relationships to other web resources. The Linked Data Platform \citep{linked-data-platform} specifies data manipulation (CRUD) operations on Linked Data resources.

In particular, in the context of the Web, Smart Components \citep{keppmann_smart_2016} paves the way to flexible and context-dependent integration. A Smart Component not only exchanges semantically defined data (through executions on its regular Web API~\cite{mariaomnivoke}) but also can adjust its program code through a so called meta API, using exactly the same interaction mechanisms enabled by the Linked Data-Fu engine \citep{harth_fly_2013}. Self-governed Components \citep{bader2017automating} utilize this functionality to independently trigger actions and achieve more robust Web service networks.

\section{Scenario: Robot Gripper Arm}
\begin{figure}[h]
\centering
\includegraphics[width=0.475\textwidth]{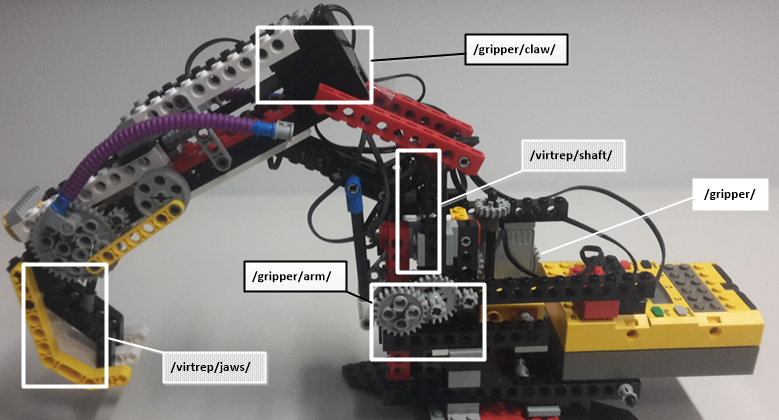}
\caption{Illustration of IoT resources (white) and \virtualrepresentation s (grey)}
\label{fig:gripper}
\end{figure}

The implementation of the here descried scenario is openly accessible. 
A Linked Data Platform server hosts the IoT representation of a demonstration gripper as an LDP RDF Source together with its components\footnote{\url{http://km.aifb.kit.edu/services/step-iot/gripper/}} (Fig. \ref{fig:gripper}). Requesting the web resource of the gripper, as well as of it sub-resources, is possible via GET requests, returning the current state of the physical object as far as observed together with some meta data. RESTful interactions with the gripper's Web API are enabled by overwriting the current resource state of the gripper's arm\footnote{\url{http://km.aifb.kit.edu/services/step-iot/gripper/arm/}} and claw\footnote{\url{http://km.aifb.kit.edu/services/step-iot/gripper/claw/}} by sending an RDF statement with the predicate \textit{saref:hasState} and the new state ('up'/'down' and 'opened'/'closed') at the object position. Example requests\footnote{\url{https://github.com/aifb/virtrep/requests/}} illustrate the use case.

Even though the gripper itself and two components are digitally accessible, these represents only a fraction of the actually installed components. In order to e.g., decide on the optimal maintenance strategy, further information about other parts is necessary too. We illustrate our strategy to bring those parts to the same integration layer by representing them as \virtualrepresentation s. Two \virtualrepresentation s are hosted on an Apache Marmotta\footnote{\url{http://marmotta.apache.org/}} server representing the shaft\footnote{\url{http://km.aifb.kit.edu/services/bader4/marmotta/ldp/ShaftContainer/shaft/}} and the jaws\footnote{\url{http://km.aifb.kit.edu/services/bader4/marmotta/ldp/JawsContainer/jaws/}}. 

\vspace{-1em}

The \virtualrepresentation \ simulates the physical object's state. To do so, the relation between the already available input data and the required features needs to be determined. Such an algorithm or heuristic is of course based on the currently most accurately available knowledge. Furthermore, in the same way the setup in the physical world changes or new information on the performance of the regarded components become known, the deployed logic of the \virtualrepresentation \ needs transparent update functionalities as well.

Let us assume that the correlation between the shaft's abrasion state and the conducted number of actions can be captured via a linear function, where the component is still perfectly fine after 10 usages but broken after 20. A straight forward heuristic would be:

\vspace{-1em}
\begin{equation}
    \hspace{-0.1cm}
    abrasion( actions ) = 
    \begin{cases}
    0 \quad , |actions| < 10 \\
    min\{\frac{1}{10} * (|actions| - 10), 1\} \ , else
    \end{cases}
    \label{eq:old}
\end{equation}

This naive function is most probably not very accurate. After better insights in the abrasion process, a more fine-grained relationship can be defined, reflected in Eq. \ref{eq:new}, hence the formerly deployed heuristic must be adjusted. We show in the following how a \virtualrepresentation \ can be initialized with Eq. \ref{eq:old} and then be easily updated with Eq. \ref{eq:new}.

\begin{equation}
    \hspace{-0.1cm}
    abrasion(actions) = 
    \begin{cases}
        0 \quad , |actions| < 10 \\
        min\{\frac{1}{1000} * (|actions| - 10)^3, 1\} \ , else
    \end{cases}
    \label{eq:new}
\end{equation}

\section{The VIRTUAL REPRESENTATION}

\sloppy{
The \virtualrepresentation \ at its core contains a) some descriptions of its physical counterpart, b) its current (simulated) state, c) the function/algorithm responsible for the calculation of derived features and a RESTful pattern to request it. A \virtualrepresentation \ is a Web resource identified by a globally unique URI. In this paper we focus on the interaction pattern and how the derivation function can be implemented, its characteristics, and how to interact with it.}

We model a \virtualrepresentation \ as a RESTful Web resource producing RDF, in particular Linked Data\footnote{source code available at \url{https://github.com/aifb/virtrep/}}. RDF, as the common data format, has the advantage of being a very mature and wide-spread standard as it serves as a corner stone of the Semantic Web Stack. Its native connection to formal knowledge representations makes RDF the data format of choice for loosely coupled environments~\cite{keppmann2016semantic}, such as the here considered shop floor, with many heterogeneous machines from different vendors.

\subsection{Configuration Function for Derived Features}

A configuration function $f$ of a \virtualrepresentation \ is a relation of derived features $y$ (e.g. the current abrasion state) by evaluating currently available input data $x$ (e.g. the number of movements). For this paper, we rely on Linked Data as our data format in order to cope the syntactical and semantic interoperability at the same time. Therefore, both input data and derived features are Linked Data triples ($x_i, y_j \in L$), with $L$ being the set of RDF triples compliant to the Linked Data principles.

As the Linked Data-Fu engine \citep{harth_fly_2013} is used to collect and process the input data $x$ in the demonstration implementation of \virtualrepresentation s, the function $f$ needs to be a two-stage process. First, a set of declarative rules in N3 syntax defines the initial data and processing steps. We denote this set according to \citep{harth_fly_2013} as a program $p$. Executing a rule of $p$ leads to either an execution of HTTP requests towards RESTful Web APIs -- responding with RDF -- or the derivation of new RDF statements locally. The second step involves a SPARQL Construct query $q$ filtering the collected data and finally generating $y$. The set of RDF triples $y$ representing a \virtualrepresentation \ is, therefore, for the demonstration implementation:
\begin{equation}
    y = f(x) = q \circ p (x)
\end{equation}

\begin{figure}[h]
\centering
\includegraphics[width=0.5\textwidth]{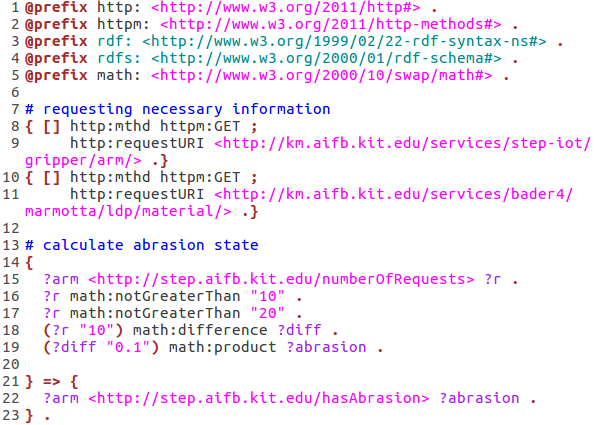}
\caption{Integrating data from external sources and deriving new information (see Eq. \ref{eq:old})}
\label{fig:program}
\end{figure}

\begin{figure}[b]
\centering
\includegraphics[width=0.47\textwidth]{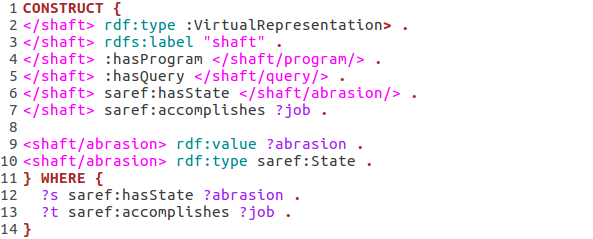}
\caption{Creating the \virtualrepresentation \ through a SPARQL Construct query}
\label{fig:query}
\end{figure}

An example program is shown in Fig. \ref{fig:program}. Lines 8 to 11 specify two HTTP request to available IoT entities as Linked Data resources. The rule from line 14 to 23 derives the new statement specifying the current abrasion state, when the condition in the body part (lines 15 to 19) of the rule is satisfied by the above requested RDF set. The query in Fig. \ref{fig:query} receives all statements from the program execution and constructs the description of the \virtualrepresentation \ as specified from line 2 to 10 in the form of a SPARQL Construct query. 

One has to note that the chosen form of the configuration function is not a general requirement for \virtualrepresentation \textit{s} but is due to the chosen engine. Different implementations can require other formats such as, for example, machine learning models or scripts. The proposed configuration method has the main advantage of being familiar with the Semantic Web and, therefore, forms a single technology stack with RDF and Linked Data.

\subsection{Interaction Model}

\begin{figure}[h]
\centering
\includegraphics[width=0.46\textwidth]{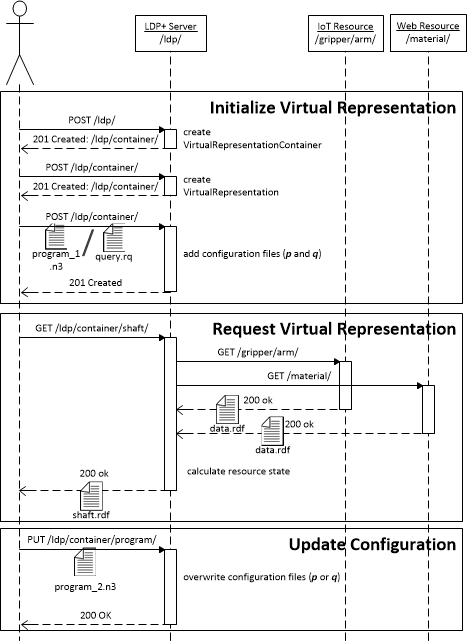}
\caption{Interaction model for \virtualrepresentation s}
\label{fig:sequence}
\end{figure}

A \virtualrepresentation \ is the combination of meta data triples together with the set of derived triples and a resource representing the function code. As this resource is also a Linked Data Platform Resource (LDP-R), it can be manipulated according to the Linked Data Platform specification. Even though the concept of \virtualrepresentation \textit{s} is conforming to the Linked Data Platform specification, \virtualrepresentation \textit{s} enhance their regular interaction model by requiring further configuration. An adaption of a Linked Data Platform server enables the proposed deployment of \virtualrepresentation \textit{s} as shown in Fig. \ref{fig:sequence}. 

The Linked Data Platform specification defines how clients can interact with LDP Web Resources in a RESTful manner. We follow the LDP specification regarding the creation, updating and deletion of Web resources. In particular, \virtualrepresentation \textit{s} and the additionally necessary configurations ($p$ and $q$) are themselves Web resources. More precisely the configurations resources are LDP Non-RDF Sources whereas the \virtualrepresentation \ itself is an LDP RDF Source. The distinction is necessary as neither the SPARQL syntax of $q$ nor Notation3 of $p$ or any other  model are necessarily RDF serializations, a requirement for an LDP RDF Source.

In general, two possible strategies can be applied to create a \virtualrepresentation \ in a RESTful manner. On the one hand, a user agent can periodically calculate the data, send it to the server, which then enables access to the latest received state of the resource. Alternatively, the server waits until a client requests the resource and then calculates the data on the fly. The first approach has advantages in an environment with unreliable network access, since at least some version of the requested data is available. Also, storing each state creates a time series and allows for historic analysis. Nevertheless, especially in the context of the IoT update rates, which are often in the range of milliseconds, this would result in creating large data volumes.
To receive a lightweight representation and also to increase the scalability of the concept~\cite{bader2017sem}, we, therefore, promote the calculation of a \virtualrepresentation 's state on the fly. An HTTP GET request on the \virtualrepresentation \ triggers the server to load both the program and the query into the engine and executes it (Fig. \ref{fig:sequence}). The dynamically created RDF statements are calculated, representing the related physical object in the same way as a regular Web resource. As we rely on the Linked Data-Fu engine the loading, compilation and execution process happens at real-time and results in a usually not notable overhead.

\begin{figure}[h]
\centering
\includegraphics[width=0.65\textwidth]{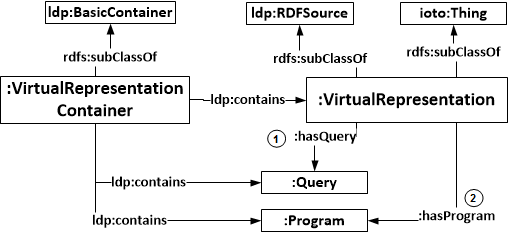}
\caption{Basic classes and relations to create a \virtualrepresentation}
\label{fig:Klassen}
\end{figure}

In order to identify the desired configuration of a \virtualrepresentation , the engine needs additional information. The relations (1) and (2) the data model (Fig. \ref{fig:Klassen}) are part of this additional information.  A naive client would expect these statements to reside at the \virtualrepresentation \ itself as both include the \virtualrepresentation \ as their subject. Unfortunately, the \virtualrepresentation \ is calculated dynamically at request time and not stored in the server's database, and, therefore, not accessible for the engine itself. Even though one could deposit the statements (1) and (2) at any other LDP resource, it is better to keep such information as close as possible to the related \virtualrepresentation . We, therefore, introduce the \virtualrepresentation \ \textit{Container}, a LDP Basic Container with the additional requirement of containing exactly one \virtualrepresentation \ together with its configuration, here the query and program resources. The engine can locate the Web resources $p$ and $q$ as the container's child resources and, therefore, combine the necessary  Web resources.

\subsection{Categorization of \virtualrepresentation s}
In REST terms, as introduced by \citep{fielding_roy_t_architectural_2000}, the proposed concept of \virtualrepresentation \textit{s} cannot be seen as a user agent. Even though the \virtualrepresentation \ can initiate request through a client connector, it is not the source of the initial request. In the same way the \textit{origin server} concept does not apply as source of the requested thing, since it is not directly connected with it. It is a general problem with the IoT that the virtual reference is not the target of the request, which in fact, is the physical object. Therefore, any IoT entity must be some kind of intermediary. \virtualrepresentation \textit{s} have server and client components as required for a \textit{proxy} but without allowing a client to choose whether to use it. In contrast, a \textit{gateway} forwards the request but acts as an origin server for the client. \virtualrepresentation \textit{s} act in that way, making them a \textit{gateway} for the digital shop floor without necessarily informing the client about this role. 

The \virtualrepresentation , as discussed in the example, is e.g. part of the 'Monitoring \& Diagnostics' module of the operations domain ('Functional Viewpoint') of the IIC Reference Architecture \citep{iic_reference_architecture} but the methodology is not restricted to it. ML trained models or the proposed program/query implementation of the configuration function are also covered by the IIC operations domain. One can locate the \virtualrepresentation \ in the underlying 'Implementation Viewpoint', since it acts as a data source or a gateway, even though, not between distinct networks but entities.

\begin{figure}[h]
\centering
\includegraphics[width=0.4\textwidth]{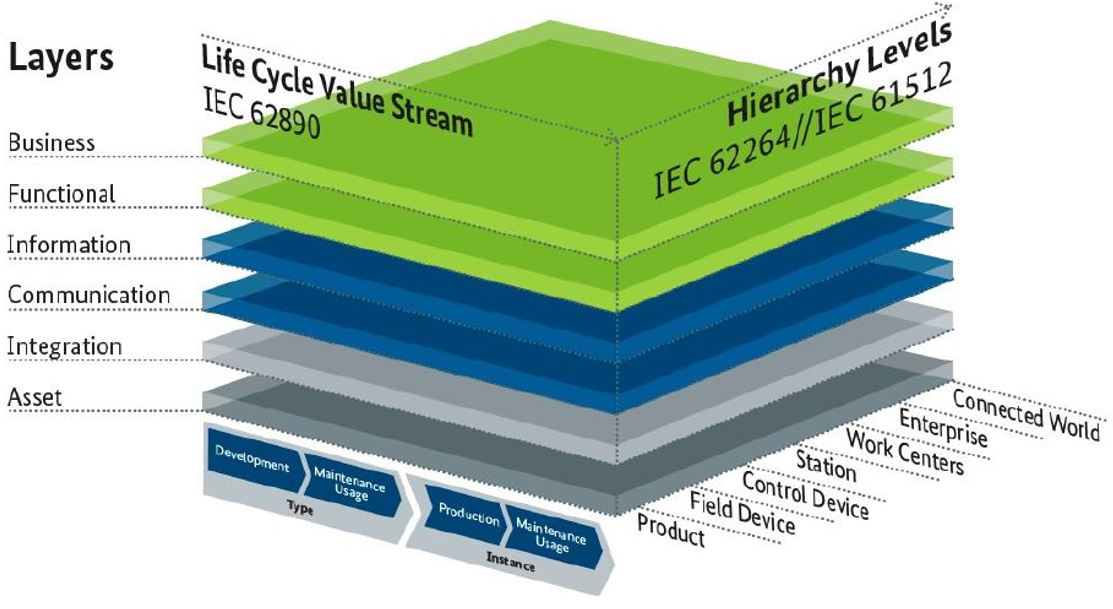}
\caption{Reference Architectural Model Industrie 4.0 according to \protect\citep{rami_fig} }
\label{fig:RAMI_ArchitectureLayers}
\end{figure}

\vspace{-1em}
The Asset Administration Shell (AAS) concept of the 'Plattform Industrie 4.0' \citep{bedenbender_industrie_2017} can act as a reference model for a \virtualrepresentation \ especially as it mentions the usage of RDF ontology concepts to clarify semantic meaning. An AAS acts as an interface between the virtual and the physical world, same as a \virtualrepresentation . Nevertheless, as discussed in \citep{bedenbender_industrie_2017}, it is not yet clear whether an AAS representing a physical objects requires a direct data connection. In the reference architecture in Fig. \ref{fig:RAMI_ArchitectureLayers} the \virtualrepresentation \ is based in the integration and communication layer, where the physical object is represented and is accessible by higher level applications \citep{rami_fig}.

\section{Conclusion and Future Work}

The concept of \virtualrepresentation \textit{s} serves as a mediator to bring unconnected physical objects quickly to the IoT. Needless to say, a simulated model cannot be as accurate as a directly observed measurement. Still, the advantage of the \virtualrepresentation \ is the low-cost deployment and the flexible ways to write the configuration into an encapsulated representation of a real-world object.

We did not focus on security-related problems and also did not discuss \virtualrepresentation \textit{s} in other IoT protocols such as MQTT, CoaP, or OPC-UA. In addition, currently only a state-based request/response interaction is implemented, in contrast to data streams and events/notifications of many IoT applications. Future steps include the consisting modeling and description of the \virtualrepresentation 's analytic logic, its interaction scheme and physical behavior in a machine-processable manner. The Hydra and openAPI vocabulary will be used to further document and explain the interface. Although we implemented an example execution environment as a HTTP Web server, the common concept is not restricted to HTTP. We aim to provide similar projects for OPC-UA and publish/subscribe protocols such as MQTT.

\virtualrepresentation \textit{s} can be one stepping stone for a faster introduction of the IoT to real-world production lines. We believe that the proposed method can reduce implementation costs and demonstrate the applicability of the approach via a simple machine demonstrator. We also describe  how \virtualrepresentation \textit{s} fit in current reference works in order to clearly specify their characteristics.

\subsubsection*{Acknowledgement}
The research and development project that forms the basis for this report is funded under project No. 01MD16015 (STEP) within the scope of the Smart Services World technology program.

\bibliographystyle{plainnat}
\bibliography{references}

\end{document}